\documentclass[12pt]{article}
\usepackage{epsfig}
\usepackage{graphicx}
\usepackage{a4}
\usepackage{latexsym}
\usepackage{cite}

\usepackage[utf8]{inputenc}

\usepackage{color}

\usepackage{pslatex}

\newcommand{\lsim}{\raisebox{-0.1cm}{$\:\:\stackrel{<}{{\scriptstyle
 \sim}}\:\: $} }
\newcommand{\beq}{\begin{equation}}
\newcommand{\eeq}{\end{equation}}
\newcommand{\bea}{\begin{eqnarray}}
\newcommand{\eea}{\end{eqnarray}}
\newcommand{\nn}{\nonumber}
\newcommand{\MSb}{$\overline{\mbox{MS}}$ }

\newcommand{\hspn}{{\hspace{-4mm}}}
\newcommand{\hspp}{{\hspace{3mm}}}

\newcommand{\as}{\alpha_{\rm s}}
\newcommand{\ar}{a_{\rm s}}

\def\z#1{{\zeta_{#1}}}
\def\z#1{{\zeta_{\:\! #1}}}

\def\zts{{\zeta_{3}^{\,2}}}

\def\ep{{\varepsilon}}

\def\xt{{(1\!-\!x)}}
\def\zt{{(1\!-\!z)}}

\def\muR {{\mu_R^{}}}
\def\muF {{\mu_F^{}}}
\def\mH{{m_H^{}}}
\def\muRs{{\mu_R^{\,2}}}
\def\muFs{{\mu_F^{\,2}}}
\def\mHs{{m_H^{\,2}}}

\def\ca{{C^{}_A}}
\def\cas{{C^{\,2}_A}}
\def\cat{{C^{\,3}_A}}
\def\caf{{C^{\,4}_A}}
\def\cf{{C^{}_F}}
\def\cfs{{C^{\, 2}_F}}
\def\cft{{C^{\, 3}_F}}
\def\cff{{C^{\, 4}_F}}
\def\nf{{n^{}_{\! f}}}
\def\nfs{{n^{\,2}_{\! f}}}
\def\nft{{n^{\,3}_{\! f}}}
\def\nff{{n^{\,4}_{\! f}}}

\def\nc{{n_c}}
\def\ncs{{n_{c}^{\,2}}}

\def\dfAAna{{\frac{d_A^{\,abcd}d_A^{\,abcd}}{n_A }}} 
\def\dfFAna{{\frac{d_F^{\,abcd}d_A^{\,abcd}}{n_A }}}
\def\dfFFna{{\frac{d_F^{\,abcd}d_F^{\,abcd}}{n_A }}}

\def\cI{{C^{}_I}}
\def\dfAInI{{\frac{d_A^{\,abcd}d_I^{\,abcd}}{n_I }}}
\def\dfRInI{{\frac{d_F^{\,abcd}d_I^{\,abcd}}{n_I }}}

\def\dfFAnc{{\frac{d_F^{\,abcd}d_A^{\,abcd}}{n_F }}}
\def\dfFFnc{{\frac{d_F^{\,abcd}d_F^{\,abcd}}{n_F }}}

\def\dfAA{d_A^{\,abcd}d_A^{\,abcd}}
\def\dfFA{d_F^{\,abcd}d_A^{\,abcd}}
\def\dfFF{d_F^{\,abcd}d_F^{\,abcd}}

\def\bz{{\beta_0^{}}}
\def\bn#1{{\beta_0^{#1}}}
\def\bone{{\beta_1^{}}}

\def\lnzt#1{{\ln^{\,#1 \!}\zt}}
\def\xiH{{\xi_H^{(4)}}}

\def\Ag4{{A_{g,4}}}
\def\Dg4{{D_{g,4}}}

\begin{document}
\setlength{\parskip}{0.15cm}
\setlength{\baselineskip}{0.52cm}

\begin{titlepage}
\renewcommand{\thefootnote}{\fnsymbol{footnote}}
\thispagestyle{empty}
\noindent
SI-HEP-2020-07 \hfill March 2020 \\
DESY 20-037 \\
LTH 1230
\vspace{1.0cm}

\begin{center}
{\bf \Large 
	Approximate four-loop QCD corrections\\[1.5mm] 
	to the Higgs-boson production cross section
}\\
  \vspace{1.25cm}
{\large
G. Das$\,$\footnote{goutam.das@uni-siegen.de}$^{a}$,
S. Moch$\,$\footnote{sven-olaf.moch@desy.de}$^{b}$
and
A. Vogt$\,$\footnote{andreas.vogt@liverpool.ac.uk}$^{c}$
   \\
}
 \vspace{1.25cm}
 {\it
   $^{a}$Theoretische Physik 1, Naturwissenschaftlich-Technische Fakult{\"a}t, Universit{\"a}t Siegen\\
   Walter-Flex-Strasse 3, D--57068 Siegen, Germany \\
   \vspace{0.25cm}
   $^{b}$II. Institut f\"ur Theoretische Physik, Universit\"at Hamburg \\
   Luruper Chaussee 149, D--22761 Hamburg, Germany \\
   \vspace{0.25cm}
   $^{c}$Department of Mathematical Sciences, University of Liverpool \\
   Liverpool L69 3BX, United Kingdom \\
 }
  \vspace{2.5cm}
  \large {\bf Abstract}
  \vspace{-0.3cm}
\end{center}
We study the soft and collinear (SV) contributions to inclusive Higgs-boson 
production in gluon-gluon fusion at four loops.
Using recent progress for the quark and gluon form factors and Mellin moments 
of splitting functions, we are able to complete the soft-gluon enhanced 
contributions exactly in the limit of a large number of colours, and to a 
sufficiently accurate numerical accuracy for QCD. 
The four-loop SV contributions increase the QCD cross section at 14 TeV by 
$2.7\%$ and $0.2\%$ for the standard choices $\muR=\mH$ and $\muR=\mH/2$ of the
renormalization scale, and reduce the scale uncertainty to below $\pm 3\%$.
As by-products, we derive the complete $\delta \xt$ term for the gluon-gluon 
splitting function at four loops and its purely Abelian contributions at
five loops, and provide a numerical result for the single pole of the 
four-loop gluon form factor in dimensional regularization.
Finally we present the closely related fourth-order coefficients $D_4^{}$ for 
the soft-gluon exponentiation of Higgs-boson and Drell-Yan lepton-pair 
production.
\end{titlepage}

%
The production of the Standard Model (SM) Higgs boson in proton-proton 
collisions and its subsequent decay are flagship measurements in run 2 of 
the Large Hadron Collider (LHC)~\cite{Sirunyan:2018koj,Aad:2019mbh}.
The main production mechanism for $pp \to H +X$ is the gluon-gluon fusion (ggF)
process. The corresponding inclusive cross section serves as a benchmark for 
the achieved accuracy, both in the LHC experiments and for theoretical research.
The radiative corrections in Quantum Chromodynamics (QCD) for the ggF process 
are large and have motivated significant efforts to improve the precision of 
the predictions.
The QCD corrections are currently known to the next-to-next-to-next-to-leading
order (N$^{3}$LO) in the effective theory for a large top-quark mass, 
$m_t \gg \mH$ \cite{Anastasiou:2015vya,Mistlberger:2018etf}, 
and to next-to-next-to-leading order (NNLO) in the full theory for Higgs-boson 
masses $\mH \lsim 2\,m_t$~\cite{Harlander:2009mq,Pak:2009dg,Harlander:2009my}.

Near the production threshold, for $z=\mHs/\hat{s}$ close to unity, where  
$\mH$ is the Higgs mass and $\sqrt{\hat{s}}\,$ the partonic center-of-mass 
energy, the QCD corrections to the ggF process are dominated by the well-known 
large logarithmic corrections. 
At $n$-th order they appear in the partonic cross section in the \MSb scheme
as plus-distributions ${\cal D}_k = [\zt^{-1}\ln^{\,k}\zt]_+$ with 
$2n-1 \ge k \ge 0$, while the virtual contributions lead to $\delta \zt$ terms.
In Mellin $N$-space, where $N$ is the conjugate variable of $z$, the threshold 
logarithms read $\ln^{\,k} N$ with $2n \ge k \ge 1$, and the virtual 
contributions lead to a constant in $N$.
The soft-virtual~(SV) approximation to the partonic ggF cross section in 
$N$-space yields reliable predictions for the total Higgs production cross 
section, as has been demonstrated with comparisons to exact fixed-order 
results up to NNLO, see, e.g., Refs.~\cite{Moch:2005ky,deFlorian:2014vta}.
In~addition, Mellin $N$-space lends itself to an all-order exponentiation of 
threshold contributions up to next-to-next-to-next-to-leading logarithmic 
(N$^3$LL) accuracy and beyond~\cite{Moch:2005ba,Das:2019btv}.

These facts motivate the derivation of approximate QCD corrections to the ggF
process at four loops in the effective theory, which can be achieved thanks to 
recent progress in the computation of QCD corrections for related quantities 
at the four-loop level.
This comprises results for specific colour contributions, including quartic
group invariants, and the planar limit of quark and gluon form factors
\cite{Lee:2016ixa,Lee:2017mip,Lee:2019zop,Henn:2019rmi,vonManteuffel:2019wbj,
Huber:2019fxe,vonManteuffel:2020vjv},
correlators of Wilson lines
\cite{Grozin:2018vdn,Henn:2019swt,Bruser:2019auj},
splitting functions for the evolution of parton distributions (PDFs)
\cite{Davies:2016jie,Moch:2017uml,Moch:2018wjh,Vogt:2018miu},
and, related, the knowledge of a low number of Mellin moments for the structure 
functions in deep-inelastic scattering (DIS), see Ref.~\cite{Ruijl:2016pkm}.

Taken together, this knowledge enables us to determine precise numerical results
for the complete SV approximation of the ggF process at four loops as well as
partial information on terms suppressed by a power $1/N$ in Mellin $N$-space
using physical evolution kernels at the same order~\cite{deFlorian:2014vta}.
The results are used to provide new predictions for the ggF cross section at
the collision energy of $14$ TeV, as planned for run 3 of the LHC.
We also present the corresponding expression for the Drell-Yan (DY) process,
$pp \to \gamma^{\,\ast} + X$, which is closely related to the ggF process 
in the threshold limit, and the N$^4$LL soft-gluon exponentiation coefficient 
$D_4^{}$ for both processes.
As by-products, we obtain a complete result for the so-called gluon virtual 
anomalous dimension, i.e. the $\delta \zt$ terms of the gluon-gluon splitting 
function at four loops in QCD, together with partial information at five loops,
and we derive a numerical result for the single pole $1/\varepsilon$ of the 
dimensionally regulated gluon form factor at four loops.

\medskip
The effective coupling of the Higgs boson to partons is described by the 
Lagrangian
\beq
\label{eq:Leff}
  {\cal L}_{\rm{eff}} \:\:=\:\: -\frac{1}{4\upsilon} \, C(\muRs) \: 
  H \, G^{\,a}_{\!\mu\nu} G_{\!a}^{\:\mu\nu} \:\: ,
\eeq
where $\upsilon \simeq 246~\rm{GeV}$ is the Higgs vacuum expectation value in
the SM and $G^{\,a}_{\!\mu\nu}$ denotes the gluon field strength tensor.
The inclusive hadronic cross section for Higgs-boson production at a 
center-of-mass energy $E_{cm}=\sqrt{S}\,$ is given in standard QCD 
factorization by
\bea
\label{eq:had}
  \sigma(S,\mHs) &\!=\!& \tau\: 
  \sum\limits_{a,b} \:\int_0^1 \frac{dx_1^{}}{x_1^{}} \;\frac{dx_2^{}}{x_2^{}} 
  \; f_{a/h_1^{}}(x_1^{},\muFs)\;f_{b/h_2^{}}(x_2^{},\muFs) 
  \int_0^1 \! dz \;\delta \Big(z - \frac{\tau}{x_1^{} x_2^{}}\Big)\, \times\,  
\nn \\[0.5mm] & & 
  \times\:\: \widetilde{\sigma}_0^{\:\!H} \: c^{\,H}_{ab}(z,\, \as(\muRs),\, 
  \mHs/\muRs,\, \mHs/\muFs) \;\; ,
\eea
where $\tau=\mHs/S$, and $\muF$, $\muR$ are the mass-factorization and 
renormalization scales, and $f_{a/h}(x,\muFs)$ the PDFs of the proton.
The expansion in the strong coupling $\as$ of the large-$m_t$ effective vertex
for the Higgs coupling to gluons is included in $\widetilde{\sigma}_0^{\:\!H}$,
viz
\beq
\label{eq:sigma0}
  \widetilde{\sigma}_0^{\:\!H} \:\: = \:\:
  \frac{\pi\, C(\muRs)^2}{8\, n^{ }_{\!A}\, \upsilon^2} 
  \quad \mbox{ with } \quad
  C(\muRs) \:\: = \:\: - \,\frac{\as(\muRs)}{3 \pi} 
  \: \Big\{ 1 \,+\, 11\: \frac{\as(\muRs)}{4 \pi} \: + \: \ldots \Big\} \;\; ,
\eeq
where $n^{}_{\!A} = (\ncs - 1)$ denotes the dimension of the adjoint
representation of the $SU(n_c)$ gauge group, and the matching coefficient 
$C(\muRs)$ is fully known up to N$^4$LO in QCD ($\nc=3$)
\cite{Chetyrkin:2016uhw}$\,$%
\footnote{See Refs.~\cite{Chetyrkin:1997un,Schroder:2005hy,Chetyrkin:2005ia} 
for previous work up to N$^3$LO in QCD.}. 
The coefficient functions $c^{\:\!H}_{ab}$ are expanded in powers of 
$\ar \equiv \as(\muRs)/(4\pi)$,
\beq
\label{eq:cexp}
 c^{\,H}_{ab}(z,\,\as(\muRs),\, \mHs/\muRs,\, \mHs/\muFs) 
  \:\:=\:\: \sum_{n=0} a_{\rm s}^{\,n} 
  c_{ab}^{\,H,(n)}(z,\,\mHs/\muRs,\,\mHs/\muFs) 
 \:\: ,
\eeq
where the leading order (LO) is
$c_{ab}^{\,H,(0)} = \delta_{ag}\,\delta_{bg}\,\delta \zt$ 
and the corrections to N$^3$LO have been computed 
in Refs.~\cite{Anastasiou:2015vya,Mistlberger:2018etf}$\,$\footnote{See 
Refs.~\cite{Dawson:1990zj,Djouadi:1991tka,Spira:1995rr,Harlander:2005rq}
and \cite{Harlander:2002wh,Anastasiou:2002yz,Ravindran:2003um} 
for previous work at lower orders in QCD.}.
For $c_{gg}^{\,H,(4)}(z)$, at N$^4$LO, seven of the eight plus-distributions 
of the SV approximation are known.  
The coefficients of ${\cal D}_k$ for $7 \geq k \geq 2$ can be inferred from 
Ref.~\cite{Moch:2005ky} and have been written down in Eq.~(16) of 
Ref.~\cite{Ravindran:2006cg} and that of ${\cal D}_{\,1}$ is fixed by the
results of Ref.~\cite{Anastasiou:2014vaa} and has been given in Eq.~(13) 
of Ref.~\cite{Ahmed:2014cla}. An approximate result for the ${\cal D}_{\,0}$ 
term has been provided before in Eqs.~(2.13) and (2.14) of 
Ref.~\cite{deFlorian:2014vta}.

Here we present a new result for the latter coefficient for a general gauge 
group. The relevant Casimir invariants for $SU(n_c)$ are $\ca = \nc$, 
$\cf = (\ncs-1)/(2 \nc)$ and 
\bea
\label{d4SUn}
  \dfAAna \:=\:
  \frac{1}{24}\: \ncs ( \ncs + 36 ) 
\; , \qquad
  \dfFAna \:=\:
  \frac{1}{48}\: n_c ( \ncs + 6 ) 
\; .
\eea
With the recent progress at four loops on the pole structure of the QCD form 
factors, on splitting functions and on Mellin moments for DIS structure 
functions, and following the same procedure as employed for DIS structure 
functions in~Ref.~\cite{Das:2019btv}, the 
${\cal D}_{\,0}$ term in $c_{gg}^{\,H,(4)}(z)$ can now be given as 
\bea
\label{eq:d0at4loops}
\lefteqn{ c_{gg}^{\,H,(4)}\Big|_{{\cal D}_0} \; = \; }
\nonumber\\
&& \nonumber      
    \caf\*\biggl(
    -\frac{40498399}{2187}
    +\frac{28613426}{729}\,\*\zeta_2
    +\frac{10995352}{81}\,\*\zeta_3
    +\frac{2598712}{81}\,\*\zeta_4
    -\frac{7252952}{27}\,\*\zeta_2\*\zeta_3
\\
&& \nonumber      
    +\frac{3411280}{9}\,\*\zeta_5
    -\frac{656216}{3}\,\*\zts
    +\frac{1019381}{9}\,\*\zeta_6
    -\frac{293488}{3}\,\*\zeta_3\*\zeta_4
    -561344\,\*\zeta_2\*\zeta_5 
    +986440\,\*\zeta_7
\\
&& \nonumber      
    +\frac{1}{12}\* ~f_{4,\, \dfFA}^{\rm q}~ 
    \biggr)
    + \dfAAna \* \biggl( - 2\,\* f_{4,\, \dfFA}^{\rm q} \,
    \biggr)
    +\cat\*\nf\*\biggl(
            \frac{17665315}{2916}
          - \frac{10870138}{729}\,\*\zeta_2
\\
&& \nonumber      
          - \frac{3234580}{81}\,\*\zeta_3
          - \frac{364960}{81}\,\*\zeta_4
          + \frac{520864}{9}\,\*\zeta_2\*\zeta_3
          - \frac{2216816}{27}\,\*\zeta_5
          + \frac{263864}{9}\,\*\zts
          - \frac{578258}{27}\,\*\zeta_6
\\
&& \nonumber      
    - ~b_{4,\, {\nf\*\cft}}^{\rm q}~
    -2\,\* b_{4,\, {\nf\*\cfs\*\ca}}^{\rm q}~
    -\frac{1}{12}\* ~b_{4,\, \dfFF}^{\rm q}~ 
    \biggr)
    +\cas\*\cf\*\nf\*\biggl(
            \frac{2798681}{486}
          - \frac{39658}{9}\,\*\zeta_2
\\
&& \nonumber
          - \frac{367508}{27}\,\*\zeta_3
          - \frac{130640}{27}\,\*\zeta_4
          + \frac{17120}{9}\,\*\zeta_2\*\zeta_3
          + \frac{21904}{9}\,\*\zeta_5
          + \frac{34064}{3}\,\*\zts
          - 988\,\*\zeta_6
    +4\,\* b_{4,\, {\nf\*\cfs\*\ca}}^{\rm q} \,
    \biggr)
\\
&& \nonumber      
    +\ca\*\cfs\*\nf\*\biggl(
          - \frac{27949}{54}
          - 632\,\*\zeta_2
          + \frac{2240}{9}\,\*\zeta_3
          + 668\,\*\zeta_4
          + \frac{1024}{3}\,\*\zeta_2\*\zeta_3
          - \frac{7744}{3}\,\*\zeta_5
\\
&& \nonumber
          - 736\,\*\zts
          + \frac{29336}{9}\,\*\zeta_6
          + 4\* ~b_{4,\, {\nf\*\cft}}^{\rm q}~
    \biggr)
    +\nf\*\dfFAna\* \biggl(
    768
    - \frac{9088}{3}\,\*\zeta_2
    + \frac{10624}{9}\,\*\zeta_3
\\
&& \nonumber
    + \frac{1600}{3}\,\*\zeta_4
    - 256\*\zeta_2\*\zeta_3
    + \frac{43520}{9}\,\*\zeta_5
    - \frac{2432}{3}\,\*\zts
    - \frac{2368}{9}\,\*\zeta_6
    + 4 \* ~b_{4,\, \dfFF}^{\rm q}~
    \biggr)
\\
&& \nonumber
    -\cas\*\nfs\*\biggl(
     \frac{1543153}{2916}
    -\frac{1171400}{729}\,\*\zeta_2
    -\frac{176624}{81}\,\*\zeta_3
    -\frac{1168}{9}\,\*\zeta_4
    +\frac{71200}{27}\,\*\zeta_2\*\zeta_3
    -\frac{34592}{9}\,\*\zeta_5
    \biggr)
\\
&& \nonumber
    -\ca\*\cf\*\nfs\*\biggl(
     \frac{155083}{243}
    -\frac{5600}{9}\,\*\zeta_2
    -\frac{4784}{9}\,\*\zeta_3
    -\frac{160}{3}\,\*\zeta_4
    +\frac{1280}{3}\,\*\zeta_2\*\zeta_3
    -32\,\*\zeta_5
    \biggr)
\\
&&
    +\ca\*\nft\*\biggl(
     \frac{10432}{2187}
    -\frac{3200}{81}\,\*\zeta_2
    -\frac{3680}{81}\,\*\zeta_3
    +\frac{112}{9}\,\*\zeta_4
    \biggr)
\:\: .
\eea
Here the term $f_{4,\, \dfFA}^{\,\rm q}$ is related to the eikonal anomalous
dimension of the four-loop quark form factor, i.e., to the single pole in 
dimensional regularization, cf.~Ref.~\cite{Das:2019btv}. 
The expressions $b_{4,\, \dfFF}^{\rm q}$, $b_{4,\, {\nf\*\cfs\*\ca}}^{\rm q}$ 
and $b_{4,\, {\nf\*\cft}}^{\rm q}$ denote the four-loop coefficients of the 
respective colour factor in the quark virtual anomalous dimension 
$B^{\,\rm q}$, i.e.,the coefficient of $\delta \zt$ in the quark-quark 
splitting function $P_{qq}$.
At $n$-th order, expanding in powers of $\ar \equiv \as(\muRs)/(4\pi)$ 
analogous to Eq.~(\ref{eq:cexp}), the flavor-diagonal splitting functions 
$P_{ii}$ admit the large-$z$ expansion~\cite{Korchemsky:1988si,Albino:2000cp} 
as
\beq
\label{xto1Lnc}
 P_{\,ii}^{\,(n-1)}(z) \:\: = \:\; A_n^{\:\!\rm i}\,{\cal D}_0 
  \,+\, B_n^{\,\rm i} \, \delta \zt \,+\, \ldots\: 
  \, ,\qquad i\,=\,{\rm q,g}
\: ,
\eeq
where the coefficients $A_n^{\:\! \rm i}$ are the well-known lightlike cusp 
anomalous dimensions.
The quantities $f_{4,\, \dfFA}^{\,\rm q}$, $b_{4,\, \dfFF}^{\rm q}$, 
$b_{4,\, {\nf\*\cfs\*\ca}}^{\rm q}$ and $b_{4,\, {\nf\*\cft}}^{\rm q}$ 
are not known analytically. They drop out in the large-$\nc$ limit of 
Eq.~(\ref{eq:d0at4loops}).
Precise numerical estimates, i.e., 
$b_{4,\, \dfFF}^{\rm q} = -143.6 \pm 0.2$, 
$b_{4,\, {\nf\*\cfs\*\ca}}^{\rm q} =  -455.247 \pm 0.005 $ and 
$b_{4,\, {\nf\*\cft}}^{\rm q} = 80.780 \pm 0.005$
have been given in Ref.~\cite{Das:2019btv} together with
$f_{4,\, \dfFA}^{\,\rm q} = -100 \pm 100$. 

\medskip
While the large-$\nc$ limit of Eq.~(\ref{eq:d0at4loops}) is exact, the above
general expression uses one assumption on the relation of quark and gluon form 
factors ${\cal F}^{\rm q}$ and ${\cal F}^{\rm g}$ in QCD in dimensional 
regularization, which, in the normalization of Eq.~(\ref{eq:cexp}), admit the 
perturbative expansion as 
\bea
  {\cal F}^{p} & = &  1 +  a_{\rm s} \left(
          - {1 \over 2\:\!\varepsilon^2} \,\* A_1^p
          - {1 \over 2\:\!\varepsilon} \,\* G_1^p
\right)
+ {\cal O}(a_{\rm s}^2)
\, \qquad  i\,=\,{\rm q,g}
\: ,
\label{eq:ffloop}
\eea
see e.g.~Refs.~\cite{Das:2019btv,Moch:2005tm} for the higher orders.

The logarithms of ${\cal F}^{\rm q}$ and ${\cal F}^{\rm g}$ contain double and 
single poles in $\varepsilon$, the former being controlled by their respective 
cusp anomalous dimensions, $A^{\rm q}$ and  $A^{\rm g}$, 
cf.~Eq.~(\ref{xto1Lnc}), which exhibit generalized Casimir scaling through 
four loops, see, e.g.~\cite{Moch:2018wjh,vonManteuffel:2020vjv}.
The single poles on the other hand, which are proportional to functions 
$G_n^p(\varepsilon)$ at $n$-th order, are controlled by the collinear anomalous
dimensions and can be converted to appropriate eikonal (Wilson line) quantities,
after separation of the virtual anomalous dimensions $B^{\:\!\rm q}$ and  
$B^{\:\!\rm g}$, cf.~Eq.~(\ref{xto1Lnc}), and terms proportional to the QCD 
$\beta$-function. In~detail (see, e.g.~\cite{Das:2019btv,Moch:2005tm}), the 
functions $G_n^p(\varepsilon)$ satisfy the following relations to five loops
\bea
\label{eq:Gff}
G_1^p &\!=\!& 2\*\left(B_1^p - \delta_{pg} \beta_0 \right)
  + f_1^p  + \varepsilon\*f_{01}^p
\, ,\nonumber \\[1mm]
G_2^p &\!=\!& 2\*\left(B_2^p - 2\delta_{pg} \beta_1 \right) 
  + (f_2^p + \beta_0\*f_{01}^p) + \varepsilon\*f_{02}^p
\, ,\nonumber \\[1mm]
G_3^p &\!=\!& 2\*\left(B_3^p - 3\delta_{pg} \beta_2 \right) 
  + (f_3^p + \beta_1\*f_{01}^p + \beta_0\*f_{02}^p) + \varepsilon\*f_{03}^p
\, ,\nonumber \\[1mm]
G_4^p &\!=\!& 2\*\left(B_4^p - 4\delta_{pg} \beta_3 \right) 
  + (f_4^p + \beta_2\*f_{01}^p + \beta_1\*f_{02}^p + \beta_0\*f_{03}^p) 
  + \varepsilon\*f_{04}^p
\, ,\nonumber \\[1mm]
G_5^p &\!=\!& 2\*\left(B_5^p - 5\delta_{pg} \beta_4 \right) 
  + (f_5^p + \beta_3\*f_{01}^p + \beta_2\*f_{02}^p + \beta_1\*f_{03}^p 
  + \beta_0\*f_{04}^p) + \varepsilon\*f_{05}^p
\, ,
\eea
where the functions $f_{0n}^p(\varepsilon)$ at $n$ loops are polynomials in
$\varepsilon$ and $\beta_n$ are the coefficients of the QCD $\beta$-function 
normalized as in Eq.~(\ref{eq:cexp}), i.e., $\beta(\ar) 
= -\beta_{0\,}^{} \ar^{\,2} - \ldots$ with $\beta_0 = 11/3\;C_A - 2/3\;\nf\,$.

The eikonal anomalous dimensions $f^{\,\rm q}$ and $f^{\,\rm g}$ of these Wilson
line quantities for quarks and gluons exhibit the same maximal non-Abelian 
colour structure as the cusp anomalous dimensions, a fact verified explicitly 
at lower fixed orders~\cite{Ravindran:2004mb,Moch:2005tm} and generalized 
in Ref.~\cite{Dixon:2008gr}.
Hence we assume here that also the expressions for $f^{\,\rm q}$ and 
$f^{\,\rm g}$ are related by generalized Casimir scaling at four loops (and 
beyond), in complete analogy to the cusp anomalous dimensions, $A^{\rm q}$ 
and $A^{\rm g}$, see also the recent work~\cite{Falcioni:2019nxk}.

This leads immediately to the expression for the full colour dependence of the 
four-loop gluon virtual anomalous dimension $B_4^{\:\!\rm g}$ as 
\bea
\nonumber
\lefteqn{ B_4^{\,\rm g} \; = \; }
\\[-2mm]
&& \nonumber      
\caf\*\biggl(
~b_{4,\, \caf}^{\rm g}~
\biggr)
+ \dfAAna \* \biggl( 
~b_{4,\, \dfAA}^{\rm g}~
\biggr)
+\nf\*\cat\*\biggl(
- \frac{8075}{108}
- \frac{6155}{54}\,\*\zeta_2
- \frac{22714}{27}\,\*\zeta_3
+ \frac{7789}{18}\,\*\zeta_4
\\
&& \nonumber      
+ \frac{1874}{9}\,\*\zeta_2\*\zeta_3
+ \frac{919}{9}\,\*\zeta_5
- \frac{1268}{3}\,\*\zts
+ \frac{1777}{54}\,\*\zeta_6
- \frac{1}{4}\* ~b_{4,\, {\nf\*\cft}}^{\rm q}~
- \frac{1}{2}\* ~b_{4,\, {\nf\*\cfs\*\ca}}^{\rm q}~
- \frac{1}{48}\* ~b_{4,\, \dfFF}^{\rm q} \,
\biggr)
\\
&& \nonumber      
+\nf\*\cas\*\cf\*\biggl(
\frac{23566}{243}
+ \frac{4198}{27}\,\*\zeta_2
+ \frac{8854}{27}\,\*\zeta_3
- \frac{27269}{27}\,\*\zeta_4
- \frac{2744}{9}\,\*\zeta_2\*\zeta_3
+ \frac{6712}{9}\,\*\zeta_5
+ \frac{1928}{3}\,\*\zts
\\
&& \nonumber      
- \frac{2879}{9}\,\*\zeta_6
+ ~b_{4,\, {\nf\*\cfs\*\ca}}^{\rm q}~
\biggr)
+\nf\*\ca\*\cfs\*\biggl(
- \frac{2723}{27}
- 162\,\*\zeta_2
+ \frac{2948}{9}\,\*\zeta_3
+ 204\*\zeta_4
+ \frac{256}{3}\,\*\zeta_2\*\zeta_3
\\
&& \nonumber      
- 912\,\*\zeta_5
- 224\,\*\zts
+ \frac{6434}{9}\,\*\zeta_6
+ ~b_{4,\, {\nf\*\cft}}^{\rm q} \,
\biggr)
+\nf\*\cft\*\biggl(
23
\biggr)
+\nf\*\dfFAna\* \biggl(
\frac{1952}{9}
- \frac{2368}{3}\,\*\zeta_2
\\
&& \nonumber      
+ \frac{1312}{3}\,\*\zeta_3
+ \frac{1016}{3}\,\*\zeta_4
+ 544\*\zeta_2\*\zeta_3
- \frac{1520}{3}\,\*\zeta_5
- \frac{1496}{9}\,\*\zeta_6
+ ~b_{4,\, \dfFF}^{\rm q} \,
\biggr)
+\nfs\*\cas\*\biggl(
\frac{1352}{81}
+ \frac{37}{27}\,\*\zeta_2
\\
&& \nonumber      
+ \frac{289}{27}\,\*\zeta_3
+ \frac{200}{27}\,\*\zeta_4
- \frac{32}{9}\,\*\zeta_2\*\zeta_3
- \frac{8}{9}\,\*\zeta_5
\biggr)
+\nfs\*\ca\*\cf\*\biggl(
\frac{3910}{243}
+ \frac{160}{9}\,\*\zeta_3
\biggr)
+\nfs\*\cfs\*\biggl(
\frac{338}{27}
- \frac{176}{9}\,\*\zeta_3
\biggr)
\\
&&
+\nfs\*\dfFFna\* \biggl(
- \frac{704}{9}
+ \frac{512}{3}\,\*\zeta_3
\biggr)
+\nft\*\ca\* \biggl(
\frac{5}{243}
\biggr)
+\nft\*\cf\* \biggl(
\frac{154}{243}
\biggr)
\label{eq:B4g}
\, ,
\eea
where the $\nft$-dependent terms agree with Ref.~\cite{Davies:2016jie}.
In addition, we have checked that a numerical fit for the $\dfFA$ term in the 
gluon-gluon splitting function~\cite{Moch:2018wjh} to the known Mellin moments 
nicely confirms the value quoted in Eq.~(\ref{eq:B4g}).  Altogether, we take 
this as strong indications on the correctness of the assumption made in the 
derivation of Eqs.~(\ref{eq:d0at4loops}) and (\ref{eq:B4g}).
The remaining unknowns can be determined numerically as 
$b_{4,\, \caf}^{\rm g} = 1098 \pm 20$ and $b_{4,\, \dfAA}^{\rm g} = -1125.6 
\pm 1.0$ from the Mellin moments $N = 2,\,4,\,6$ and 8 obtained via DIS 
structure functions, see~Ref.~\cite{Vogt:2018miu}.

We also note that the purely Abelian (QED) contributions in $B_4^{\:\!\rm g}$ 
coincide with the respective terms in the four-loop QCD $\beta$-function
\cite{vanRitbergen:1997va,Czakon:2004bu}. 
This concerns the colour factors $\nf\*\cft$, $\nfs\*\cfs$, $\nft\*\cf$ and 
$\dfFF$ and is a direct consequence of the generalized Casimir scaling of 
$f^{\,\rm q}$ and $f^{\,\rm g}$, which implies that $f_4^{\rm g}$ must have 
only non-Abelian colour factors (terms proportional to $\ca$, $\dfFA$ or 
$\dfAA$).
This reproduces a pattern already observed for $B_n^{\:\!\rm g}$ 
up to third order, $n \leq 3$, see \cite{Vogt:2004mw}, for all terms 
$n_{\! f}^{\,k\:} C_F^{\:n-k}$ with $1 \,\leq\, k \,\leq\, n\!-\!1$.
At four loops, the colour factors $\nf\*\cft$ and $\dfFF$ are unique 
in the single pole $1/\varepsilon$ of the gluon form factor, 
cf.~in $G_4^g$ in Eq.~(\ref{eq:Gff}), and therefore, can be related 
directly to those in $\beta_3$ and, hence, $B_4^{\:\!\rm g}$.
The other two colour factors in $G_4^g$, $\nfs\*\cfs$ and $\nft\*\cf$, do also 
receive contributions from lower orders. For instance $\nfs\*\cfs$ terms are 
generated from $\beta_1\*f_{02}^g$, but cancel in extraction of 
$B_4^{\:\!\rm g}$ from $G_4^g$.

With the help of Eqs.~(\ref{eq:Gff}) and (\ref{eq:B4g}) we obtain the single 
pole in $\varepsilon$ in the gluon form factor ${\cal F}_4^{g}$ at four loops 
as 
\bea
\nonumber
\lefteqn{ {\cal F}_4^{g}\biggr|_{1/\varepsilon} \; = \; }
\\[-2mm]
&& \nonumber      
\caf\*\biggl(
- \frac{746918615}{104976}
+ \frac{595199}{216}\,\*\zeta_2
+ \frac{8305667}{1458}\,\*\zeta_3
+ \frac{975575}{972}\,\*\zeta_4
+ \frac{39811}{81}\,\*\zeta_2\*\zeta_3
- \frac{781411}{405}\,\*\zeta_5
\\
&& \nonumber      
+ \frac{41335}{54}\,\*\zeta_3\*\zeta_4
- \frac{272338}{81}\,\*\zts
- \frac{739783}{144}\,\*\zeta_6
- \frac{14629}{45}\,\*\zeta_2\*\zeta_5
+ \frac{563669}{126}\,\*\zeta_7
+\frac{1}{192}\* ~f_{4,\, \dfFA}^{\rm q}~ 
\\
&& \nonumber
-\frac{1}{4}\* ~b_{4,\, \caf}^{\rm g}~
\biggr)
+ \dfAAna \* \biggl( 
- \frac{80}{9}
+ \frac{704}{3}\,\*\zeta_3
-\frac{1}{8}\* ~f_{4,\, \dfFA}^{\rm q}~ 
-\frac{1}{4}\* ~b_{4,\, \dfAA}^{\rm g}~
\biggr)
+ {\cal O}(\nf)
\\
&=&
\caf\*\Bigl(
- 1084.7 \pm 5.5
\biggr)
+ \dfAAna \* \Bigl( 
567.3 \pm 12.8
\biggr)
+ \nf \mbox{ terms}
\label{eq:ff4loopsinglepole}
\; ,
\eea
where all $\nf$-dependent contributions have been given analytically in 
Ref.~\cite{vonManteuffel:2020vjv}.

The observed relation between the gluon virtual anomalous dimension 
$B_n^{\:\!\rm g}$ and the corresponding coefficient of the $\beta$-function for purely Abelian terms leads to new predictions at five 
loops.
Using the expression for the $\beta$-function for a general gauge group at 
five loops ~\cite{Herzog:2017ohr,Luthe:2017ttg,Chetyrkin:2017bjc} (the QCD 
result was obtained before in Ref.~\cite{Baikov:2016tgj}), one deduces for 
the splitting function $P_{gg}$ in Eq.~(\ref{xto1Lnc})
\bea
\nonumber
\lefteqn{ B_5^{\:\!\rm g} \; = \; }
\\[-2mm] && \nonumber      
         \nf\* \cff\* \biggl( 
         - {4157 \over 12}  
         - 64 \,\* \zeta_3 \biggr)
       + \nfs \* \cft\* \biggl( 
         - {2509 \over 18}  
         - {536 \over 3} \,\* \zeta_3 
         + {1160 \over 3} \,\* \zeta_5 \biggr)
\\ && \nonumber
       + \nfs \* \cf\* \dfFFna \* \biggl( 
           {4160 \over 3}  
         + {5120 \over 3} \,\* \zeta_3 
         - {12800 \over 3} \,\* \zeta_5 \biggr)
       + \nft \* \cfs\* \biggl( 
         - {4961 \over 324}  
         + {952 \over 27} \,\* \zeta_3 
         - {44 \over 3} \,\* \zeta_4 \biggr)
\\ && \nonumber
       + \nft\* \dfFFna \* \biggl( 
           {1760 \over 9}  
         - {1312 \over 3} \,\* \zeta_3
         + 128\,\* \zeta_4 
         + {640 \over 3} \,\* \zeta_5 \biggr)
       + \nff \* \cf\* \biggl( 
         - {107 \over 486}  
         - {8 \over 27} \,\* \zeta_3 \biggr)
\\[1mm] && \nonumber      
\label{eq:B5g}
  \,+\, \mbox{ terms with }\; \ca,\, \dfAA,\, \dfFA
\: .
\eea
Predictions for the purely Abelian part of the gluon form factor at five loops 
are possible, e.g., for the $\nf\cff$ in ${\cal F}_5^{g}$, which can be read 
off from Eq.~(\ref{eq:B5g}) using $G_5^g$ in Eq.~(\ref{eq:Gff}), while finite 
terms of ${\cal F}^{g}$ at lower orders are needed for other colour 
structures.

\medskip
Beyond the SV approximation, predictions for the ggF cross section are possible
using physical evolution kernels
\cite{deFlorian:2014vta,Moch:2009hr,Soar:2009yh}.
In $z$-space, this concerns terms enhanced as $\ln^{\,k}\zt$ with $2n-1 \ge k 
\ge 1$ at $n$-th order, or equivalently power suppressed contributions 
$(\ln^{\,k} N)/N$ in Mellin $N$-space.
Such next-to-leading power threshold effects have also been studied in
Refs.~\cite{Grunberg:2009yi,Grunberg:2009vs,Almasy:2010wn,Almasy:2015dyv,%
DelDuca:2017twk,Bahjat-Abbas:2019fqa,Beneke:2019mua,Beneke:2019oqx}.
At N$^4$LO, these subleading terms in $c_{gg}^{\,H,(4)}(z)$ can be obtained
from the physical evolution kernel $K_{gg}$, which one can define by 
re-expressing Eq.~(\ref{eq:had}) as dimensionless `structure functions' 
${\cal F}_{ab}$, i.e.,
\beq \textstyle
  \sigma(S,\mHs) \:\: = \:\: \sum_{a,b} \, \widetilde{\sigma}_0^{\,H} \: 
  {\cal F}_{ab} \:\: .
\eeq
The kernel $K_{gg}$ and its perturbative expansion for a scale choice 
$\muF = \mH$ are then given in terms of the splitting function $P_{gg}$, the 
$\beta$-function and the gluon coefficient function $c^H_{gg}$ by
\bea
\label{eq:physkern}
  \frac{d}{d\ln \mHs}\:{\cal F}_{gg} &\!=\!&
  \bigg\{ 2 P_{gg}(a_s) + \beta(a_s) \: \frac{d c^{\,H}_{gg}(a_s)}{d a_s} 
  \,\otimes\,
  \left(c^{\,H}_{gg}(a_s)\right)^{-1} \bigg\} \otimes {\cal F}_{gg}
\nn \\[-1mm]
 &\!\equiv\!& 
  K_{gg}\otimes{\cal F}_{gg} \;\equiv\; 
  \sum_{\ell=0} \, a_s^{\,\ell+1} K_{gg}^{\,(\ell)}\otimes {\cal F}_{gg}
\eea
where $\otimes$ denotes the usual Mellin convolution and $\beta(\ar)$ the 
$\beta$-function as defined below Eq.~(\ref{eq:Gff}).

The key feature of the kernel $K_{gg}$ is its single-logarithmic enhancement. 
In $z$-space, this implies at $n$-th order that all terms $\ln^{\,k}\zt$ with 
$2n-1 \ge k \ge n+1$ have to cancel in Eq.~(\ref{eq:physkern}) to all orders 
in $\zt$, which leads to predictions for coefficient function $c^{\,H}_{gg}$, 
cf.~Ref.~\cite{deFlorian:2014vta}.
In Mellin $N$-space the leading large-$N\,$ logarithms of the sub-dominant 
$N^{\,-1}$ contributions in $K_{gg}$ take the simple form,
\bea
\label{eq:xiH}
  \left. K_{gg}^{(1)}\right|_{N^{\,-1}} &\!=\!& 
  - \,\left( 8\,\beta_0\, \ca + 32\,\cas \right)\, \ln N \:+\; {\cal O}(1)
\:\: , 
\nn \\[1mm]
  \left. K_{gg}^{(2)}\right|_{N^{\,-1}} &\!=\!&
  - \,\left( 16\,\beta_0^2\, \ca + 112\,\beta_0\, \cas \right)\, \ln^{\,2\!} N 
  \:+\; {\cal O}( \ln N)
\:\: , 
\nn \\[1mm]
  \left. K_{gg}^{(3)}\right|_{N^{\,-1}} &\!=\!&
  - \,\left( 32\,\beta_0^3\, \ca + \frac{896}{3}\, \beta_0^2\, \cas \right)\,
  \ln^{\,3\!} N \:+\; {\cal O}( \ln^{\,2\!} N)
\:\: , 
\nn \\[1mm]
  \left. K_{gg}^{(4)}\right|_{N^{\,-1}} &\!=\!&
  - \,\left( 64\,\beta_0^4\, \ca + \xiH\, \beta_0^3\, \cas \right)\,
  \ln^{\,4\!} N \:+\; {\cal O}( \ln^{\,3\!} N)
\,,
\eea
where the first three lines follow from the known coefficient functions 
$c_{gg}^{\:\!H,(n)}$ up to N$^3$LO.
The expression for $K_{gg}^{(4)}$ contains an unknown coefficient $\xiH$, 
to be determined 
at N$^4$LO by explicit computations (the corresponding coefficient for DIS
has been obtained in Refs.~\cite{Grunberg:2009vs,Almasy:2010wn})$\,$\footnote{
We note that the pattern of the ratios of the lower order coefficients is 
$112/32=7/2$ and $(896/3)/112=8/3$. A~generalization of this pattern leads 
to an estimate of $\xiH=670 \pm 300$ with a conservative numerical 
uncertainty, which has a sub-percent effect on the $N^{-1} \ln^4 N$ 
coefficient in Eq.~(\ref{eq:c4gN}) below.}.
Eq.~(\ref{eq:xiH}) predicts the following next-to-leading power threshold
terms in the four-loop gluon coefficient function $c_{gg}^{\:\!H,(4)}$ for 
the ggF process at the scale $\,\muR \,=\, \muF \,=\, \mH$,
\bea
\label{eq:c4z}
  c_{gg}^{\,H,(4)}(z) &\!\!=\!\!&
  c_{gg}^{\,H,(4)}(z)\Big|_{\rm SV}
  \:-\: \frac{4096}{3}\: \* \caf\: \* \lnzt7
  + \left\{
    \frac{19712}{3}\: \* \caf + \frac{3584}{3}\: \* \cat\, \* \bz
  \right\}\, \* \lnzt6
\nn \\[1mm] && \mbox{\hspn} 
  + \:\left\{
      \left( -\, 2240 + 23552\, \* \z2 \right) \* \caf
    - \frac{64576}{9} \* \cat\,\* \bz
    - \frac{1024}{3}\: \* \cas\, \* \bn2
  \right\}\, \* \lnzt5
\nn \\[1mm] && \mbox{\hspn} 
  + \:\left\{
   \left( \frac{80384}{9} - 81520\* \z2 - \frac{313600}{3}\* \z3 \right)\* \caf
  + \left(\frac{104128}{9} - \frac{47408}{3}\* \z2 \right) \* \cat\,\* \bz
\right.
\\ && \mbox{\hspp} 
\left.
    + \bigg(1856 + \frac{1}{4}\* \xiH \bigg) \* \cas\, \* \bn2
    + 32\: \* \ca\, \* \bn3
    + \frac{640}{3}\: \* \cas\, \* \bone
  \right\}\, \* \lnzt4
\nn 
  \,+\, {\cal O}\left( \ln^{\,3\!}\zt \right) \quad
\eea
where $c_{gg}^{\:\!H,(4)}(z)\big|_{\rm SV}$ denotes the $z$-space SV 
approximation at N$^4$LO as discussion above.

\medskip
We are now ready to assemble the results of 
Refs.~\cite{Ravindran:2006cg,Ahmed:2014cla,deFlorian:2014vta} and
Eqs.~(\ref{eq:d0at4loops}) and (\ref{eq:xiH}) for the inclusive ggF process 
at N$^4$LO.  The resulting threshold expansion of $c_{gg}^{\:\!H,(4)}$ in 
$N$-space reads 
\bea
\label{eq:c4gN}
  \kappa_4^{}\, c_{gg}^{\:\!H,(4)}(N) \!&\!\simeq\!&
    0.55296\, \ln^{\,8\!}N
  + 3.96654\, \ln^{\,7\!}N
  + 21.2587\, \ln^{\,6\!}N
  + 62.2985\, \ln^{\,5\!}N
\nn \\[1mm] & & \mbox{\hspn\hspn}
  + 150.141\, \ln^{\,4\!}N
  + 212.443\, \ln^{\,3\!}N
  + 255.911\, \ln^{\,2\!}N
  + \bigl( 128.78 \pm 0.11 \bigr) \ln N
  \,+\, \kappa_4^{}\, g_{_0,4}
\nn \\ & & \mbox{\hspn\hspn}
  + \: N^{\,-1} \!\left\{
    2.21184\,\ln^{\,7\!}N
  + 19.6890\,\ln^{\,6\!}N
  + 93.0439 \ln^{\,5\!}N
\right.
\nn \\[-1mm] & & \mbox{\hspn}
\left.
  + ( 256.454 + 132.25 \,\kappa_4^{}\,\xiH ) \ln^{\,4\!}N
  + {\cal O} ( \ln^{\,3\!}N )
  \!\right\}
\eea
with $\kappa_4^{} = 1/25000 \simeq 1/(4\pi)^4$, which approximately converts 
the coefficients to an expansion in~$\as$.
We have inserted the QCD values of the group factors in Eq.~(\ref{d4SUn}) and 
above, used the physical value of $\nf = 5$ light flavors at scales of order 
$\mHs$, and truncated all exact numbers to six decimals.
The quoted uncertainty in the coefficient of $\ln N$ stems entirely from 
$f_{4,\, \dfFA}^{\,\rm q}$, as the uncertainties in the values of 
$b_{4,\, \dfFF}^{\rm q}$, $b_{4,\, {\nf\*\cfs\*\ca}}^{\rm q}$ and 
$b_{4,\, {\nf\*\cft}}^{\rm q}$ are completely negligible.
The constant-$N$ contribution $g_{0,4}^{}$ has been estimated in Ref.~%
\cite{deFlorian:2014vta} by three Pad\'e approximants which yield a fairly 
wide spread of values suggesting $\kappa_4^{}\,g_{0,4}^{} = 65\pm 65$. 

\begin{figure}[ht]
  \centerline{
     \includegraphics[width=8.0cm]{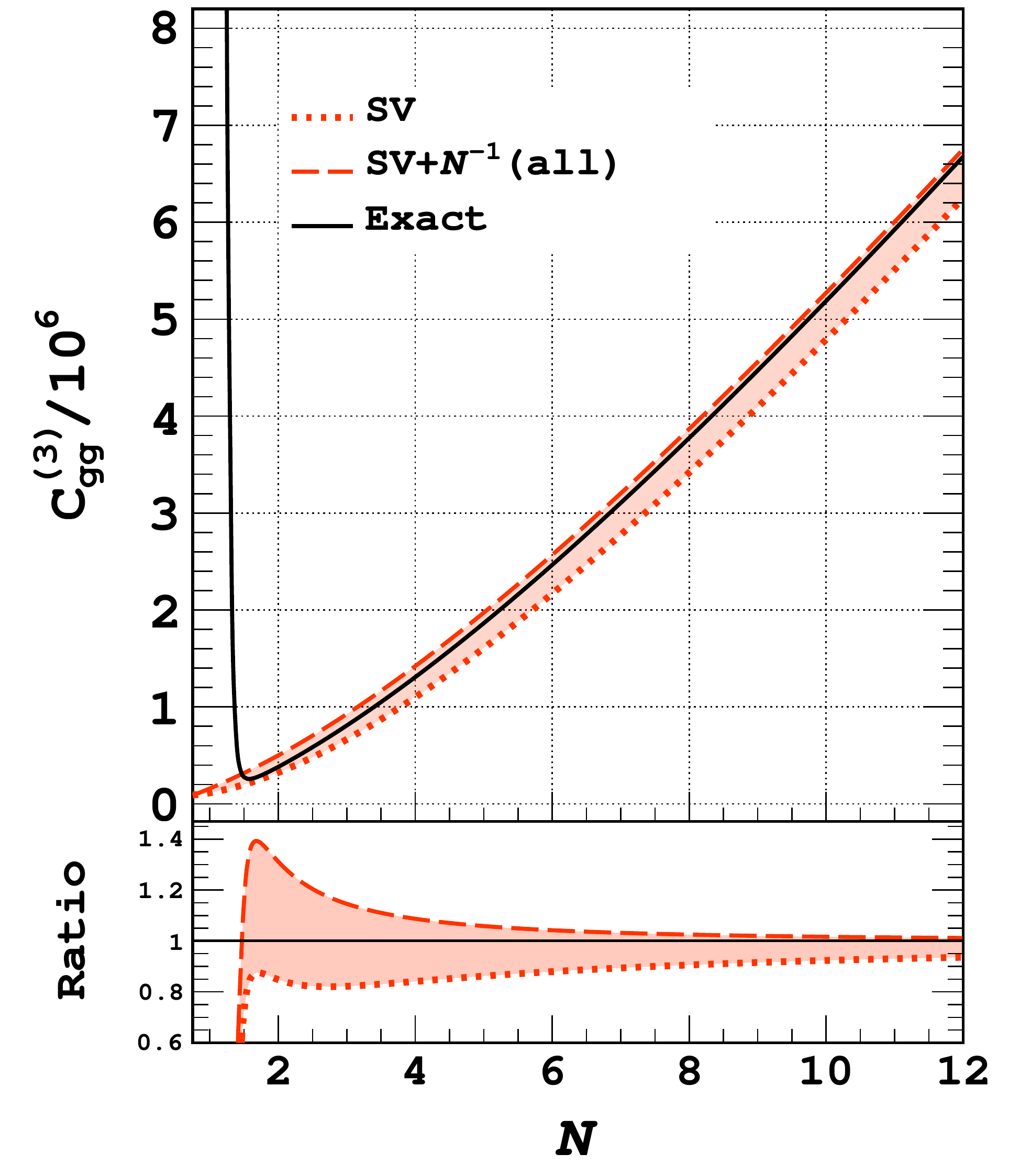}
     \includegraphics[width=8.0cm]{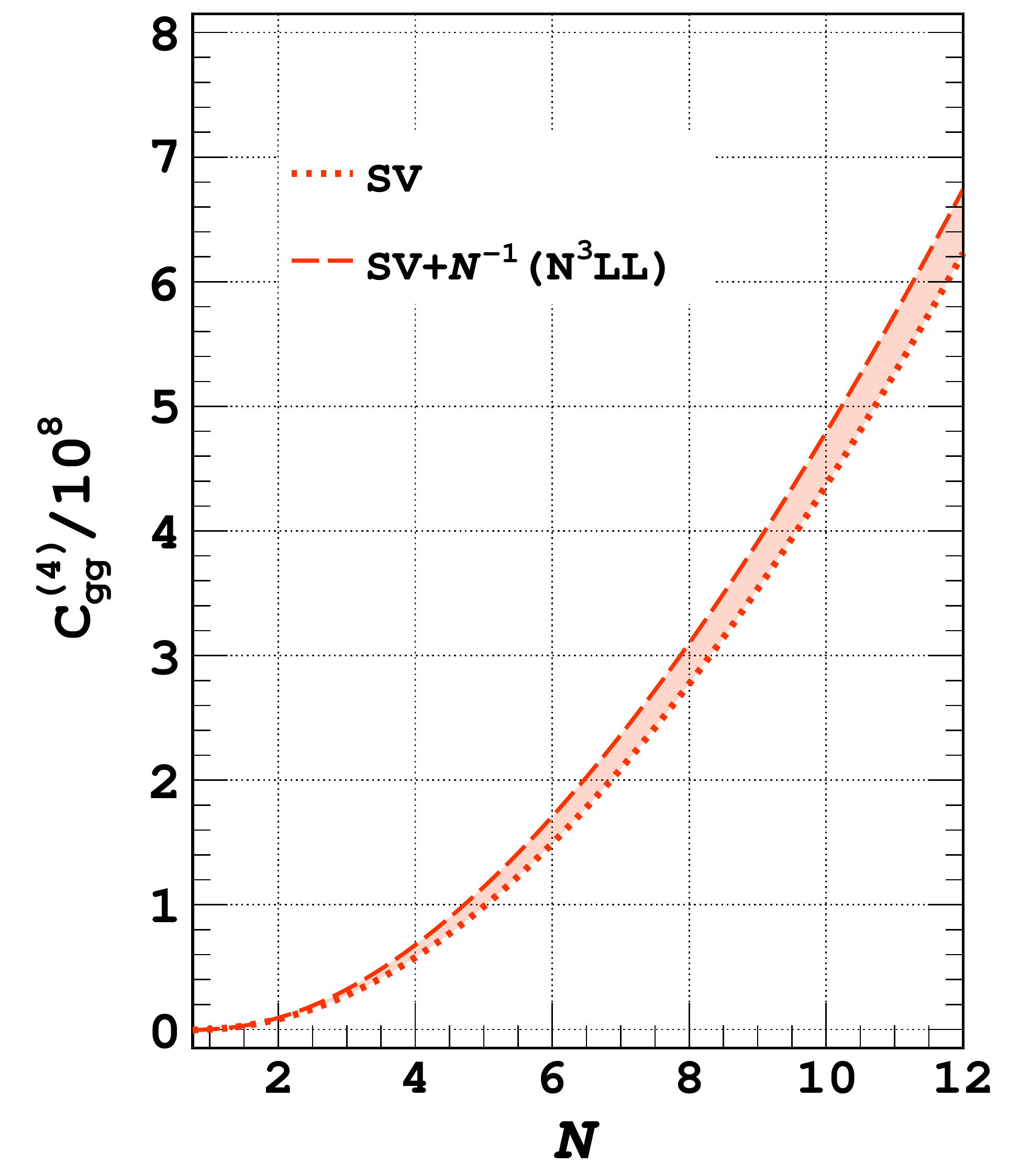}
  }
  \vspace*{-2mm}
  \caption{  \label{fig:mellin_var}
  \small{%
  Soft-virtual (SV) and SV $+N^{\,-1}$ approximations to the N$^3$LO (left) 
  and N$^4$LO (right) ggF coefficient functions. The N$^3$LO curves, where
  the $N^{\,-1}$ contributions are complete, are compared to the exact result.
  At N$^4$LO the highest four $N^{\,-1}$ logarithms are includes as given in
  Eq.~(\ref{eq:c4gN}).}
  }
\end{figure}

The N$^4$LO coefficient function in the SV approximation, together with the 
sub-dominant $N^{\,-1}$ contributions in Mellin $N$-space, can be expected
to provide a reliable approximation of the exact result.
As pointed out earlier, the exact Mellin $N$-space result at lower orders 
resides inside the band spanned by the SV and SV$+N^{\,-1}$ terms at 
moderately large $N$.
This is shown in Fig.~\ref{fig:mellin_var} (left) at N$^3$LO (for corresponding
NLO and NNLO plots see Fig.~1 of Ref.~\cite{deFlorian:2014vta}).
The exact coefficient functions differ from the approximation based on the 
SV$+N^{\,-1}$ terms by $0.44\%$ at NLO, $0.83\%$ at N$^2$LO and $1.15\%$ at 
N$^3$LO at $N=12$.  
For smaller $N$ values the difference between the exact results and the 
approximation based on the SV$+N^{\,-1}$ terms is larger, however they always
remain inside the SV and SV$+N^{\,-1}$ band.
At N$^4$LO, see Fig.~\ref{fig:mellin_var} (right), the SV approximation 
of Eq.~(\ref{eq:c4gN}) is shown including the $N$-independent constant 
$g_{04}^{}$ and the known $1/N$ terms as specified in Eq.~(\ref{eq:c4gN}).
%

The predictions for the ggF cross sections at the collision energy of $14$ TeV 
use a Higgs mass $\mH=125$ GeV, an on-shell top quark mass $m_t = 172.5$ GeV, 
$\nf=5$ active quark flavors and the PDF sets ABMP16~\cite{Alekhin:2017kpj} 
and MMHT2014~\cite{Harland-Lang:2014zoa} using the {\tt lhapdf} 
\cite{Buckley:2014ana} interface. 
The PDF sets and as well as the value of the strong coupling constant $\as$ 
corresponding to the respective PDF set are taken order-independent at NNLO 
throughout.
The prefactor $C(\muRs)$ in Eq.~(\ref{eq:sigma0}) is improved with the full 
top-mass dependence of the Born cross section.
The results up to N$^3$LO are computed with the program {\it iHixs}
\cite{Dulat:2018rbf} which directly provides the cross sections in this
rescaled effective field theory. 

The residual uncertainty in the SV approximation of the ggF cross section due 
to the coefficients $f_{4,\, \dfFA}^{\,\rm q}$ in Eq.~(\ref{eq:d0at4loops}) 
and $g_{04}^{}$ in Eq.~(\ref{eq:c4gN}), which are currently least constrained, 
is practically negligible:
the 100\% error on $g_{04}^{}$ leads to an uncertainty of $0.3\%$ for the 
cross section, while the 100\% error on $f_{4,\, \dfFA}^{\,\rm q}$ is 
completely negligible.
The virtual anomalous dimension $B_4^{\:\!g}$ only appears in the scale-%
dependent terms at N$^4$LO and its contribution vanishes for the central scale 
choice $\muR=\muF=m_H$.  For the scale setting $\muR=\muF=m_H/2$ a change 
below 0.002\% is observed for the cross-section at $14$ TeV LHC due to the 
numerical uncertainty in $B_4^g$.
Thus, precise predictions at N$^4$LO are now possible for all relevant 
kinematics and scale choices.

\begin{figure}[ht]
  \centerline{
    \includegraphics[width=8.0cm]{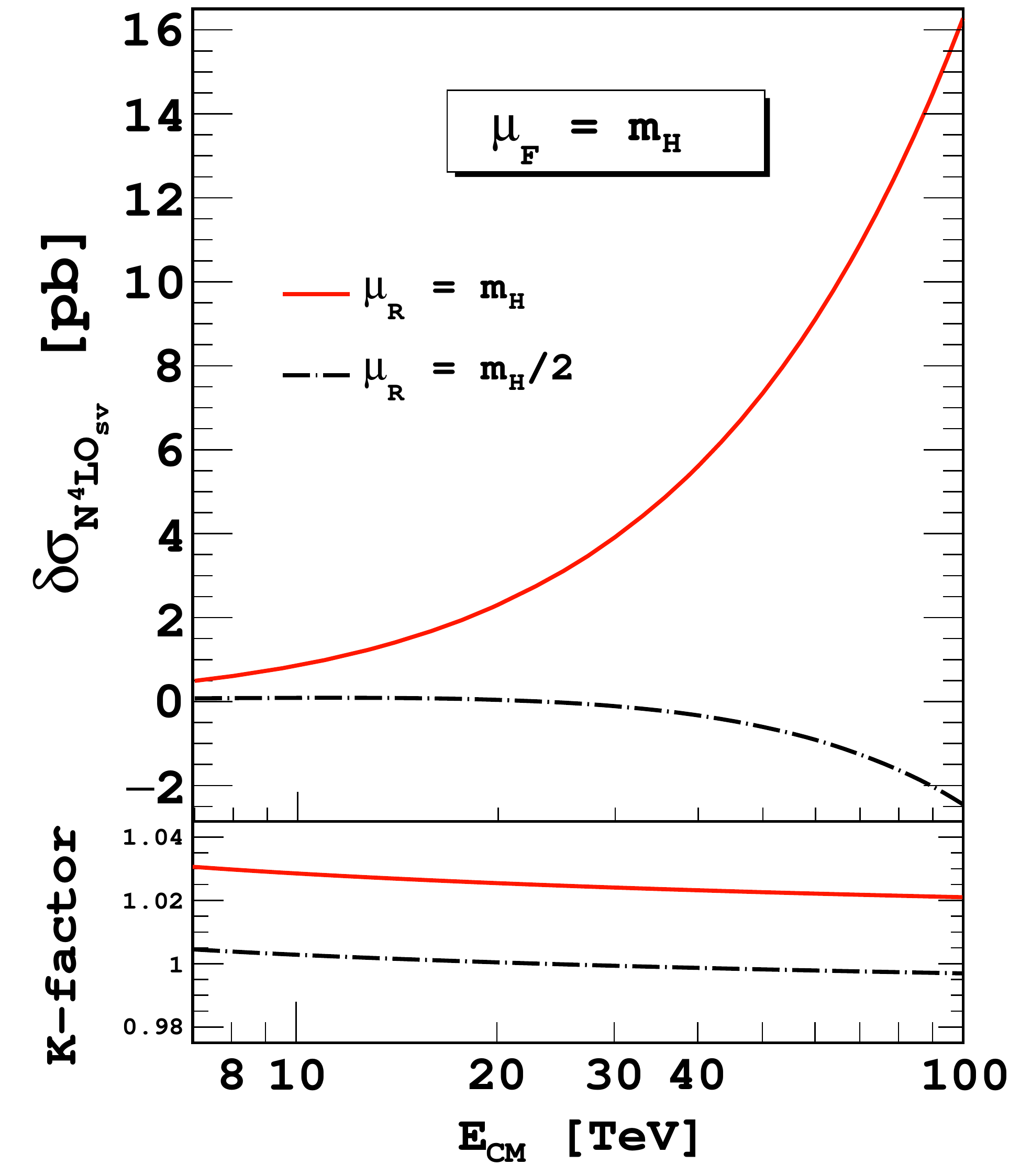}
    \includegraphics[width=8.0cm]{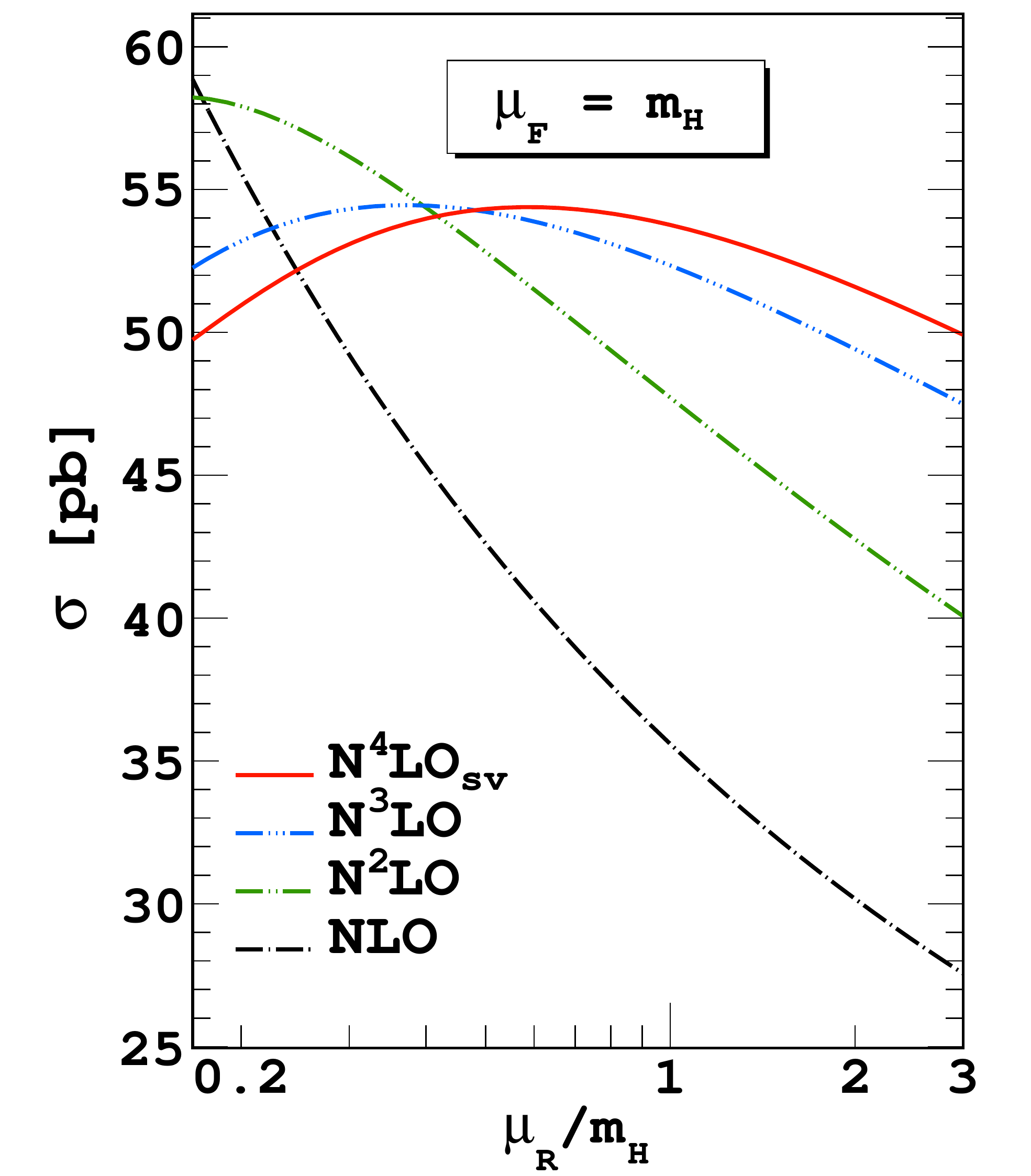}
  }
  \vspace{-2mm}
  \caption{\small{%
Left panel:
The contribution of the SV approximation at N$^4$LO to the ggF cross section
in proton-proton collisions as a function of the center-of-mass energy 
$E_{\rm CM}$ for $\muF=m_H$ at two renormalization scales, $\muR=\mH$ and 
$\muR=\mH/2$, using the MMHT2014 PDFs~\cite{Harland-Lang:2014zoa}. 
The corresponding $K$-factors with respect to N$^3$LO are shown in the lower 
panel.
Right panel:
The ggF cross section up to N$^4$LO with a variation of the renormalization 
scale $\muR$ for the LHC at $14$ TeV using the same PDFs and $\muF=\mH$.
}}
\label{fig:ecm_mu_var}
\end{figure}
The impact of the SV corrections at N$^4$LO is shown in 
Fig.~\ref{fig:ecm_mu_var} (left) in a range of center-of-mass energies for two 
different choices of renormalization scale, $\muR=\mH$ and $\muR=\mH/2$, 
keeping always $\muF=m_H$.
The corresponding $K$-factors, defined as the ratio of the SV corrections at 
N$^4$LO over the exact N$^3$LO result, are displayed in the lower panel.
The SV correction at N$^4$LO  increases the cross-section at the LHC with $14$ 
TeV by  $1.41$ pb  for the scale $\muR=\mH$ and by $0.08$ pb for the scale 
$\muR=\mH/2$.
The $K$-factor shows little dependence on the collision energy in the entire 
range of center-of-mass energies displayed in Fig.~\ref{fig:ecm_mu_var} (left).
For the scale choice $\muR=m_H/2$, which is closer to the point of minimal 
sensitivity, the effect of higher order corrections is indeed small, leading 
to a $K$-factor close to unity, cf. also~Ref.~\cite{deFlorian:2014vta}.
At the scale $\muR=m_H$ a $K$-factor of $1.027$ is obtained for the LHC at 
$14$ TeV.

In Fig.~\ref{fig:ecm_mu_var} (right) we show the dependence of the cross 
section for the $E_{\rm CM} = 14$ TeV on the renormalization scale $\muR$.
The $\muR$ dependence indeed decreases order by order in perturbation theory 
up to N$^4$LO. 
In the range $\mH/4 \leq \muR \leq 2m_H$ the effect of the $\muR$ variation
decreases drastically 
from $\pm 27\%$ at NLO and $\pm 14\%$ at NNLO to $\pm 5\%$ at N$^3$LO,
while it amounts to less than $\pm 3\%$ at N$^4$LO. Also
here the factorization scale is kept fixed, at $\muF=\mH$, since beyond NNLO
only flavour non-singlet results have published for the QCD splitting functions
$P_{\:\!ik}(z)$ \cite{Davies:2016jie,Moch:2017uml,Herzog:2018kwj}, and 
PDFs fits have been limited to NNLO so far.

The uncertainty in the predicted ggF cross sections due to the truncation of 
the perturbation series is now, at N$^4$LO, smaller than that due to
the use of different sets of PDFs and corresponding different values of the 
strong coupling constant $\as$. 
For $\sqrt{S}=14$ TeV, $\mH = 125$ GeV, the central scale $\muR=\mH$, and 
including the PDF uncertainties at N$^3$LO, one obtains 
\bea
\label{eq:higgsXS}
\nonumber
\sigma\bigr|_{\rm{N}^3\rm{LO}} 
  &\!=\!& 49.6 \pm 0.5~{\rm pb}
\, , 
\qquad
\sigma\bigr|_{\rm{N}^4\rm{LO}} 
  \!=\! 50.8~{\rm pb}
\, , 
\qquad\quad {\mbox{ABMP16}}
\, , 
\\[1mm]
\sigma\bigr|_{\rm{N}^3\rm{LO}} 
  &\!=\!& 52.3\pm 0.8~{\rm pb} 
\, , 
\qquad
\sigma\bigr|_{\rm{N}^4\rm{LO}} 
  \!=\! 53.8~{\rm pb} 
\, , 
\qquad\quad {\mbox{MMHT2014}}
  \; ,
\eea
where the spread in predictions is due to different values of the strong 
coupling constant at NNLO corresponding to the different PDF sets used, i.e., 
$\as(M_Z)=0.1147$ for ABMP16 and $\as(M_Z)=0.1180$ for MMHT2014,
and due to different gluon PDFs in the relevant kinematic range. 
These are consequences of different choices for the theoretical framework 
and assumptions on parameters used in the respective global fits, see 
Ref.~\cite{Accardi:2016ndt},
which lead to systematic shifts that are often significantly larger 
than the PDF and $\as(M_Z)$ uncertainties associated to individual PDF
sets.

\medskip
Due to the universality of threshold dynamics for colourless final states in 
hadronic collisions, relevant formulae for the Drell-Yan process, $pp \to 
\gamma^{\,\ast} + X$, can be easily obtained from the above considerations, 
using Eq.~(\ref{eq:had}) with the replacement $\widetilde{\sigma}_0^{\,H} \: 
c^{\,H}_{ab} \to \widetilde{\sigma}_0^{\gamma^{\,\ast}}\: c^{\,\rm DY}_{ab}$
with
\beq
\label{eq:DYsigma0}
\widetilde{\sigma}_0^{\gamma^{\,\ast}} \:\: = \:\:
  4 \:\!\pi\, \alpha^2/ (3 Q^2\,\nc) 
\;\; .
\eeq
Here $\alpha$ is the fine-structure constant of QED and $Q^2$ the virtuality 
of the produced photon $\gamma^{\,\ast}$.
The coefficient functions $c^{\,\rm DY}_{ab}$ enjoy a perturbative expansive 
analogous to Eq.~(\ref{eq:cexp}) with the leading order normalization 
$c_{ab}^{\,\rm DY,(0)} = \delta_{aq}\,\delta_{b{\bar q}}\,\delta \zt$.
The coefficients of ${\cal D}_k$ for $7 \geq k \geq 2$ 
of the four-loop term $c_{q{\bar q}}^{\,\rm DY,(4)}$ can be found in Eq.~(6)
of Ref.~\cite{Ravindran:2006cg} and that of ${\cal D}_{\,1}$ in Eq.~(14) 
of Ref.~\cite{Ahmed:2014cla}.

We are now in the position to present the four-loop ${\cal D}_{\,0}$ term for 
the DY process. It is given~by
\bea
\label{eq:DYd0at4loops}
\lefteqn{ c_{q{\bar q}}^{\,\rm DY,(4)}\Big|_{{\cal D}_0} \; = \; }
\nonumber\\
&& \nonumber      
\cff\*\biggl(
32704\,\*\zeta_3
+113152\,\*\zeta_2\*\zeta_3
-196608\,\*\zeta_5
-15360\,\*\zts
-491520\,\*\zeta_2\*\zeta_5
-195584\,\*\zeta_3\*\zeta_4
\\
&& \nonumber
+983040\,\*\zeta_7
\biggr)
+\cft\*\ca\*\biggl(
-\frac{206444}{27}
-\frac{32740}{9}\,\*\zeta_2
-\frac{746878}{9}\,\*\zeta_3
+\frac{146768}{9}\,\*\zeta_4
-\frac{1011088}{9}\,\*\zeta_2\*\zeta_3
\\
&& \nonumber
+274432\,\*\zeta_5
-\frac{484192}{3}\,\*\zts
+\frac{356048}{3}\,\*\zeta_6
-73728\,\*\zeta_2\*\zeta_5
+76000\,\*\zeta_3\*\zeta_4
\biggr)
\\
&& \nonumber
+\cfs\*\cas\*\biggl(
\frac{15086188}{729}
-\frac{12535492}{729}\,\*\zeta_2
+\frac{3043898}{81}\,\*\zeta_3
+\frac{2522080}{81}\,\*\zeta_4
-\frac{2968640}{27}\,\*\zeta_2\*\zeta_3
\\
&& \nonumber
+\frac{1046528}{9}\,\*\zeta_5
-\frac{82592}{3}\,\*\zts
-\frac{30184}{3}\,\*\zeta_6
+3072\*\zeta_2\*\zeta_5
+\frac{60944}{3}\,\*\zeta_3\*\zeta_4
\biggr)
+\cf\*\cat\*\biggl(
-\frac{28325071}{2187}
\\
&& \nonumber
+\frac{5761670}{243}\,\*\zeta_2
+\frac{867584}{27}\,\*\zeta_3
-\frac{150632}{9}\,\*\zeta_4
-\frac{119624}{9}\,\*\zeta_2\*\zeta_3
-\frac{49840}{9}\,\*\zeta_5
-\frac{4664}{3}\,\*\zts
+\frac{41789}{9}\,\*\zeta_6
\\
&& \nonumber
+832\,\*\zeta_2\*\zeta_5
+1440\,\*\zeta_3\*\zeta_4
+3400\,\*\zeta_7
+\frac{1}{12}\* ~f_{4,\, \dfFA}^{\rm q}~ 
\biggr)
+ \dfFAnc\* \biggl(-2\* ~f_{4,\, \dfFA}^{\rm q}~ \biggr)
\\
&& \nonumber
+\cft\*\nf\*\biggl(
- \frac{80221}{54}
- \frac{25744}{27}\,\*\zeta_2
+ \frac{95936}{9}\,\*\zeta_3
- \frac{11492}{9}\,\*\zeta_4
+ \frac{189824}{9}\,\*\zeta_2\*\zeta_3
- \frac{130624}{3}\,\*\zeta_5
\\
&& \nonumber
+ \frac{106336}{3}\,\*\zts
- \frac{160840}{9}\,\*\zeta_6
+ 4\* ~b_{4,\, {\nf\*\cft}}^{\rm q}~
\biggr)
+\cfs\*\ca\*\nf\*\biggl(
- \frac{955285}{1458}
+ \frac{3057110}{729}\,\*\zeta_2
- \frac{1222648}{81}\,\*\zeta_3
\\
&& \nonumber
- \frac{1261168}{81}\,\*\zeta_4
+ \frac{306400}{9}\,\*\zeta_2\*\zeta_3
- 37616\*\zeta_5
+ \frac{11728}{3}\,\*\zts
- \frac{164}{3}\,\*\zeta_6
+ 4\* ~b_{4,\, {\nf\*\cfs\*\ca}}^{\rm q}~
\biggr)
\\
&& \nonumber
+\cf\*\cas\*\nf\*\biggl(
\frac{10761379}{2916}
- \frac{2418814}{243}\,\*\zeta_2
- \frac{948884}{81}\,\*\zeta_3
+ \frac{213280}{27}\,\*\zeta_4
+ \frac{28064}{9}\,\*\zeta_2\*\zeta_3
- \frac{29552}{27}\,\*\zeta_5
\\
&& \nonumber
- \frac{9736}{9}\,\*\zts
- \frac{32930}{27}\,\*\zeta_6
- \frac{1}{12}\* ~b_{4,\, \dfFF}^{\rm q}~ 
- 2\* ~b_{4,\, {\nf\*\cfs\*\ca}}^{\rm q}~
- ~b_{4,\, {\nf\*\cft}}^{\rm q}~
\biggr)
\\
&& \nonumber
+ \nf\*\dfFFnc\* \biggl(
768
- \frac{9088}{3}\,\*\zeta_2
+ \frac{10624}{9}\,\*\zeta_3
+ \frac{1600}{3}\,\*\zeta_4
- 256\,\*\zeta_2\*\zeta_3
+ \frac{43520}{9}\,\*\zeta_5
- \frac{2432}{3}\,\*\zts
\\
&& \nonumber
- \frac{2368}{9}\,\*\zeta_6          
+4\* ~b_{4,\, \dfFF}^{\rm q}~ \biggr)
+\cfs\*\nfs\*\biggl(
- \frac{142769}{729}
- \frac{99184}{729}\,\*\zeta_2
+ \frac{113456}{81}\,\*\zeta_3
+ \frac{23200}{27}\,\*\zeta_4
\\
&& \nonumber
- \frac{79360}{27}\,\*\zeta_2\*\zeta_3
+ \frac{33056}{9}\,\*\zeta_5
\biggr)
+\cf\*\ca\*\nfs\*\biggl(
- \frac{898033}{2916}
+ \frac{293528}{243}\,\*\zeta_2
+ \frac{87280}{81}\,\*\zeta_3
- \frac{1744}{3}\,\*\zeta_4
\\
&&
- \frac{608}{9}\,\*\zeta_2\*\zeta_3
+ \frac{608}{3}\,\*\zeta_5
\biggr)
+\cf\*\nft\*\biggl(
\frac{10432}{2187}
-\frac{3200}{81}\,\*\zeta_2
-\frac{3680}{81}\,\*\zeta_3
+\frac{112}{9}\,\*\zeta_4
\biggr)
\:\: ,
\eea
where the quartic Casimirs are normalized by the dimension of the fundamental
representation of the $SU(n_c)$ gauge group, $n^{}_{\!F} = \nc$. 
As Eq.~(\ref{eq:d0at4loops}), this result is exact in the large-$\nc$ limit 
and has an amply sufficient numerical accuracy for all phenomenological 
applications in QCD.

The next-to-leading power threshold terms for the DY process can also be 
derived with the help of the corresponding physical evolution kernel $K_{qq}$, 
which exhibits the same simple form in Mellin $N$-space as Eq.~(\ref{eq:xiH})
for the leading large-$N\,$ logarithms of the $N^{\,-1}$ contributions with 
the obvious replacement $C_A \to C_F$,
see Ref.~\cite{deFlorian:2014vta} for further details.

Finally, by combining our new result (\ref{eq:d0at4loops}) with Eq.~(2.13) 
in Ref.~\cite{deFlorian:2014vta}, and proceeding analogously with its Drell-Yan
counterpart (\ref{eq:DYd0at4loops}), we can derive the four-loop coefficient 
$D_4$ for the soft-gluon exponentiation of inclusive Higgs production via ggF 
and DY lepton-pair production.
The two results are, as expected, related by generalized Casimir scaling 
\cite{Moch:2018wjh} which reduces to standard `numerical' $C_A / C_F$ 
lower-order Casimir scaling in the their exact large-$\nc$ limit:
\bea
\label{eq:D4HplusDY}
\lefteqn{ D_4 \; = }
\nonumber\\
&& \nonumber      
       \cI\*\cat \* \biggl(  
       - \frac{28325071}{2187} 
       + \frac{5761670}{243}\,\*\zeta_2 
       + \frac{867584}{27}\,\*\zeta_3 
       - \frac{150632}{9}\,\*\zeta_4 
       - \frac{119624}{9}\,\*\zeta_2\*\zeta_3 
       - \frac{49840}{9}\,\*\zeta_5 
\\
&& \nonumber
       - \frac{4664}{3}\,\*\zts 
       + \frac{41789}{9}\,\*\zeta_6 
       + 832\,\*\zeta_2\*\zeta_5 
       + 1440\,\*\zeta_3\*\zeta_4 
       + 3400\,\*\zeta_7 
       +\frac{1}{12}\*\, f_{4,\, \dfFA}^{\rm q}~ 
       \biggr)
\\
&& \nonumber
       - \dfAInI\* \biggl( 2\*\,f_{4,\, \dfFA}^{\rm q}~ \biggr)
       + \nf\*\cI\*\cas \* \biggl( 
         \frac{10761379}{2916} 
       - \frac{2418814}{243}\,\*\zeta_2 
       - \frac{948884}{81}\,\*\zeta_3 
       + \frac{213280}{27}\,\*\zeta_4 
\\
&& \nonumber
       + \frac{28064}{9}\,\*\zeta_2\*\zeta_3 
       - \frac{29552}{27}\,\*\zeta_5 
       - \frac{9736}{9}\,\*\zts 
       - \frac{32930}{27}\,\*\zeta_6 
       - \frac{1}{12}\*\, b_{4,\, \dfFF}^{\rm q}~ 
       - 2\*\, b_{4,\, {\nf\*\cfs\*\ca}}^{\rm q}~
       - \,b_{4,\, {\nf\*\cft}}^{\rm q}~
       \biggr)
\\
&& \nonumber
       + \nf\*\cI\*\cf\*\ca \* \biggl( 
         \frac{2149049}{486} 
         - \frac{56222}{27}\,\*\zeta_2 
         - \frac{8932}{9}\,\*\zeta_3 
         - \frac{113360}{27}\,\*\zeta_4 
         + \frac{3808}{9}\,\*\zeta_2\*\zeta_3 
         + \frac{21904}{9}\,\*\zeta_5 
\\
&& \nonumber
         + \frac{6800}{3}\,\*\zts 
         - 1436\,\*\zeta_6 
         + 4\* ~b_{4,\, {\nf\*\cfs\*\ca}}^{\rm q}~
         \biggr)
       + \nf\*\cI\*\cfs \* \biggl(  
       - \frac{27949}{54}
       - 632\,\*\zeta_2 
       + \frac{2240}{9}\,\*\zeta_3 
       + 668\,\*\zeta_4 
\\
&& \nonumber
       + \frac{1024}{3}\,\*\zeta_2\*\zeta_3 
       - \frac{7744}{3}\,\*\zeta_5 
       - 736\,\*\zts 
       + \frac{29336}{9}\,\*\zeta_6 
       + 4\*\,b_{4,\, {\nf\*\cft}}^{\rm q}~
       \biggr)
       + \nf\*\dfRInI\* \biggl(
       768 
       - \frac{9088}{3}\,\*\zeta_2 
\\
&& \nonumber
       + \frac{10624}{9}\,\*\zeta_3 
       + \frac{1600}{3}\,\*\zeta_4 
       - 256\,\*\zeta_2\*\zeta_3
       + \frac{43520}{9}\,\*\zeta_5 
       - \frac{2432}{3}\,\*\zts 
       - \frac{2368}{9}\,\*\zeta_6 
       + 4\* ~b_{4,\, \dfFF}^{\rm q}~ 
       \biggr)
\\
&& \nonumber
       + \nfs\*\cI\*\ca \* \biggl(  
       - \frac{898033}{2916} 
       + \frac{293528}{243}\,\*\zeta_2 
       + \frac{87280}{81}\,\*\zeta_3 
       - \frac{1744}{3}\,\*\zeta_4 
       - \frac{608}{9}\,\*\zeta_2\*\zeta_3 
       + \frac{608}{3}\,\*\zeta_5 
       \biggr)
\\ 
&& \nonumber
       + \cI\*\nfs\*\cf \* \biggl(  
       - \frac{110059}{243} 
       + 384\,\*\zeta_2
       + \frac{10768}{27}\,\*\zeta_3 
       + \frac{160}{3}\,\*\zeta_4 
       - 256\,\*\zeta_2\*\zeta_3 
       + 32\,\*\zeta_5 
       \biggr)
\\
&&
       + \cI\*\nft\* \biggl( 
         \frac{10432}{2187} 
       - \frac{3200}{81}\,\*\zeta_2 
       - \frac{3680}{81}\,\*\zeta_3 
       + \frac{112}{9}\,\*\zeta_4 
       \biggr)
\qquad
\eea
with 
$\cI= \cf$, $d_I^{\,abcd} = d_F^{\,abcd}$ and $n_I= n_F$ for the DY case, and 
$\cI= \ca$, $d_I^{\,abcd} = d_A^{\,abcd}$ and $n_I= n_A$ for Higgs production.
The lower-order coefficients can be found in Eqs.~(33) - (35) of
Ref.~\cite{Moch:2005ky}.

With this result, and the approximate values of Ref.~\cite{Herzog:2018kwj}
for the small effect of the five-loop cusp anomalous dimensions $A_5$, all
ingredients are now available for extending the soft-gluon exponentiation
to the next-to-next-to-next-to-next-to-leading logarithmic (N$^4$LL) accuracy.
The corresponding function $g_5^{}$ can be inferred from the DIS result in
Eq.~(2.9) of Ref.~\cite{Das:2019btv} as described below Eq.~(3.6) of
Ref.~\cite{Moch:2005ba}.

\medskip
Using recent progress on related fourth-order quantities, we have been able 
to determine the final soft-gluon enhanced contribution (\ref{eq:d0at4loops}) 
to the N$^4$LO coefficient function for inclusive Higgs-boson production in 
gluon-gluon fusion in the heavy-top limit, and the corresponding result 
(\ref{eq:DYd0at4loops}) for the Drell-Yan process $pp \to \gamma^{\,\ast} + X$.
These results also fix the respective N$^4$LL coefficients $D_4$ for the 
soft-gluon exponentiation (\ref{eq:D4HplusDY}) which are related by the same 
fourth-order generalization of the well-known Casimir scaling observed before 
in the cusp anomalous dimensions, now completely known at this order
\cite{Henn:2019swt}.
Our results are exact in the limit of a large number of flavours $\nc$.
Their colour-factor decomposition in full QCD involves a few quantities which
are known only numerically at this point. The resulting uncertainties are
practically negligible as can be seen from the $\ln N$ coefficient in 
Eq.~(\ref{eq:c4gN}) above.

We have employed the latter Mellin $N$-space results to add the N$^4$LO 
soft + virtual (SV) corrections to the known complete N$^3$LO results 
\cite{Anastasiou:2015vya,Mistlberger:2018etf} for the LHC at 14 TeV.
With the effect of the only uncomputed quantity, the soft-gluon coefficient 
$g_{04}^{}$ for this process, being well below 1\%, we find that the cross 
sections are enhanced by 2.7\% for the scale choice $\muR = \mH$, while the 
results are almost unchanged for $\muR = 0.5\,\mH$. 
It should be noted that these values refer to the not entirely realistic case 
of an order-independent $\as$-value and  PDFs at $\mu = \mH$.
The renormalization-scale variation, estimated using the interval
$0.25 \mH\leq \muR \leq 2m_H$, is reduced from about 5\% at N$^3$LO to less
than 3\% at N$^4$LO. Based on similar calculations at lower orders, we 
definitely expect that difference between the present $N$-space SV 
approximation and the complete N$^4$LO coefficient function will amount to 
well below 1\% of the total cross section.

As by-products of our analysis, we have derived the expression (\ref{eq:B4g})
for the four-loop gluon virtual anomalous dimension (and determined the
corresponding purely Abelian contributions at five loops), and provided a
sub-percent accurate value (\ref{eq:ff4loopsinglepole}) for the hitherto 
unknown $1/\ep$ coefficient of the matter-independent contribution to the 
four-loop gluon form factor for which the $\nf$-terms have been recently 
computed in Ref.~\cite{vonManteuffel:2020vjv}.

%
\subsection*{Acknowledgements}
G.D. thanks M.C. Kumar and V. Ravindran for useful discussions. 
S.M. acknowledges useful communication with B. Mistlberger on 
Ref.~\cite{Mistlberger:2018etf}.
The algebraic computations have been done with the latest version of the
symbolic manipulation system {\sc Form}~\cite{Vermaseren:2000nd,Ruijl:2017dtg}.
This work has been supported by the {\it Deutsche Forschungsgemeinschaft} (DFG) 
under grant number MO~1801/2-1, and by the COST Action CA16201
PARTICLEFACE supported by {\it European Cooperation in Science and Technology} 
(COST).
The research of G.D. is supported by the DFG within the Collaborative Research
Center TRR 257 (``Particle Physics Phenomenology after the Higgs Discovery'').

\end{document}